\documentclass{amsart}

\usepackage{enumitem}
\usepackage[all]{xy}
\usepackage[margin=1.00in, marginparwidth=1.5cm, marginparsep=0.5cm]{geometry}
 \usepackage{booktabs}
 \usepackage{siunitx}
\usepackage[table]{xcolor} 
\usepackage{mathtools}
\usepackage{bbold}
\usepackage{placeins}
\usepackage{amssymb}
\usepackage{array}
\usepackage{xcolor}
\usepackage{mathrsfs}
\usepackage{physics}
\usepackage{amsthm}
\usepackage{amsmath,amsfonts,graphicx,mathdots,cite,wrapfig}
\usepackage[colorlinks,linkcolor=blue,citecolor=blue,pagebackref,hypertexnames=false, breaklinks]{hyperref}
\usepackage{cite}
\usepackage{mwe}
\usepackage{caption}
\usepackage{caption} 
\usepackage{subfigure} 
\usepackage{marginnote}

\newtheorem{theorem}{Theorem}[section]
\newtheorem{lemma}[theorem]{Lemma}
\newtheorem{proposition}[theorem]{Proposition}

\theoremstyle{definition}

\theoremstyle{remark}

\numberwithin{equation}{section}

\DeclareMathOperator{\complex}{\mathbb{C}}

\begin{document}



\title{Quantitative approach to Grover's quantum walk on graphs}


\author{Gamal Mograby}
\address{Department of Mathematics, University of Maryland, College Park, MD 20742, USA}
\email{gmograby@umd.edu}

\author{Radhakrishnan Balu}
\address{Radhakrishnan Balu, Department of Mathematics \& Norbert Wiener Center for
	Harmonic Analysis and Applications, University of Maryland, College Park, MD 20742, USA}
\email{rbalu@umd.edu}

\author{Kasso A. Okoudjou}
\address{Kasso A. Okoudjou, Department of Mathematics, Tufts University, Medford, MA 02155, USA}
\email{kasso.okoudjou@tufts.edu}

\author{Alexander Teplyaev}
\address{Alexander Teplyaev, Department of Mathematics, University of Connecticut, Storrs, CT 06269, USA}
\email{alexander.teplyaev@uconn.edu}


\date{\today}


\begin{abstract}

In this paper we study Grover's search algorithm  focusing on continuous-time quantum walk on graphs. We propose an alternative optimization approach to Grover's algorithm on graphs that can be summarized as follows: instead of finding specific graph topologies convenient for the related quantum walk, we fix the graph topology and vary the underlying graph Laplacians. As a result, we search for the most appropriate analytical structure on graphs endowed with fixed topologies yielding better search outcomes. We discuss strategies to investigate the optimality of Grover's algorithm and provide an example with an easy tunable graph Laplacian to investigate our ideas.

\end{abstract}

\maketitle

\tableofcontents

\section{Introduction}

The theory of quantum algorithms has been an active area of study over the last three decades, see  \cite{ChildsVanDam2010,ChuangBook2000, Santha2008, Mosca2008} and references therein.
In several applications,  quantum algorithms have been shown to outperform their classical counterparts and hence leading to a speedup in performance \cite{Shor1994,Grover1997}. In this paper we revisit  Grover's search algorithm  \cite{Grover1997,Ambainis2004,AaronsonAmbainis2005,Tulsi2008,Dowling2006,Childs2002,ShenviKempe2003,AmbainisKempe2005}, focusing on the continuous-time quantum walk approach developed by \cite{FarhiGutmann1998,ChildsGoldstone2004}. The Childs-Goldstone approach is very versatile and can be realized on quantum systems of different geometrical or topological arrangements. This  feature appeared in \cite{SpectraPST2021Mograby,PQST2020Mograby,SolitonsGradedGraphs2021Mograby}  where it was established that a certain class of fractal-type graphs demonstrates favorable topological properties to implement perfect quantum state transfer. The feature is also present in \cite{AgliariEtal2010}, where Grover's search algorithm was analyzed on databases with different topological arrangements. In this latter case, the (analytical and numerical) investigations of quantum walk on several graphs such as  the dual Sierpinski gaskets, T-fractals, and hierarchical structures like Cayley trees, illustrate the dependency of Grover's algorithm on the topological structure of these graphs.

In this paper we propose an alternative optimization approach to Grover's algorithm on graphs. In particular, instead of finding specific graph topologies convenient for the related quantum walk, we fix the graph topology and vary the underlying graph Laplacians. As a result, we search for the most appropriate analytical structure on graphs endowed with fixed topologies yielding better outcomes in Grover's search algorithm. 
To describe our approach's main ideas, we first introduce some basic terminology and notation. We perform a Grover's search on a database modeled by a finite (possibly directed) graph $G=(V,E)$. Let $\{p(x,y)\}_{(x,y) \in E}$ be a sequence of weights assigned to the edges, where we regard the edge $(x,y)$ as pointing from the vertex $x$ to $y$ and $p(x,y)$ as a transition probability of a random walker from $x$ to $y$. We impose the following conditions
\begin{align}
\label{eq:CodForTransProb}
\begin{cases}
\  (x,y) \in E \ \Leftrightarrow  \ 0<p(x,y) \leq1 \\
\  (x,y) \notin E \ \Leftrightarrow  \ p(x,y) =0 \\
\ \sum_{y :(x,y) \in E} p(x,y) = 1, \ \forall \ x \in V.
\end{cases}
\end{align}
and associate with such a sequence a \textit{probabilistic graph Laplacian} on $G$, defined by
\begin{align}
\label{eq:probLapOnGp}
 \Delta_G f (x)= f(x) \  - \sum_{y :(x,y) \in E} p(x,y) f(y),
\end{align}
We assume there exists a Hilbert space  $\big(\mathcal{H}, \langle .,.\rangle \big)$ such that $\Delta_{G}$ is self-adjoint, 
\begin{align*}
 \langle \Delta_{G} \psi,\phi \rangle = \langle \psi, \Delta_{G} \phi \rangle , \quad \phi , \psi \in \mathcal{H} = \big\{ f:V \to \complex \big \}.
\end{align*}
We refer to \cite{BaluMograbyOkoudjouTeplyaevJacobi2021} for more details, and for examples of such Hilbert spaces on certain graphs. We associate each item in the database with a vertex $x \in V$ or equivalently the corresponding normalized Dirac function
\begin{align*}
e_{x}:= {\delta_{x}}/ \sqrt{\langle \delta_{x}, \delta_{x} \rangle}, \quad \quad
\delta_x(y) :=
\begin{cases}
1, \quad x=y \\
0, \quad x \neq y
\end{cases}
\end{align*}
and denote the target vertex in Grover's  search algorithm by $w \in V$. Note that $\mu: V \to (0,\infty)$, $\mu(x):=\langle \delta_{x}, \delta_{x} \rangle$ defines a measure on the  set of vertices $V$. The \textit{volume} of the graph $G$ is then given by
\begin{align}
vol(G) := \sum_{x \in V} \mu(x), \quad \mu(x)=\langle \delta_{x}, \delta_{x} \rangle
\end{align}
To perform the search, one needs a driving Hamiltonian $H$ of the quantum system. In this work, we use (see \cite{AgliariEtal2010,ChildsGoldstone2004})
\begin{align}
\label{eq:GroverHamilton}
\begin{cases}
 \  H_{\gamma}: = \gamma \Delta_{G} - V_w , \\
  \  V_w f := \langle e_w, f \rangle  e_w, \quad f : V(G) \to \complex
 \end{cases}
\end{align}
where $\gamma$ is a tunable parameter in $(0,\infty)$. The potential operator $V_w$ is also called the \textit{oracle Hamiltonian}. As the initial state of the search we choose the ground state of $\Delta_{G}$
\begin{align}
\label{eq:initial}
s := \frac{1}{\sqrt{vol(G)}} \sum_{x \in V(G)} \delta_{x} 
\end{align}
and the goal is to evolve $s$ continuously to the target state $e_w$. The success probability of finding the target vertex $w$ at the time $t$ is then given by
\begin{align}
\label{eq:SuccessProb}
\pi_{w}^{\gamma}(t) := \big\vert \langle e_w, \exp(-iH_{\gamma}t)  s \rangle \big\vert^2.
\end{align}
Our main contribution is summarized as follows. Rather than investigating a family of Hamiltonians $\{H_{\gamma}\}_{\gamma \in (0,\infty)}$ on graphs of different topologies, we  fix the (topology on) graph $G$  and vary the transition probabilities (\ref{eq:CodForTransProb}) of a random walker on $G$. By doing so, we are effectively varying $\Delta_G$ in (\ref{eq:probLapOnGp}). As such we are led  to the following question: ``can we construct examples for which this approach improves Grover's search outcomes?''. In section \ref{sec:SHL}, we provide  an example with an easy tunable parameter to answering this question.

The rest of the paper is organized as follows: in section \ref{sec:GroverFiniteGraph}, we will discuss strategies to investigate the optimality of Grover's search algorithm. We recall that  the algorithm is implemented using a family of Hamiltonians $\{H_{\gamma}\}_{\gamma \in (0,\infty)}$ for which one is led to determine in a systematic manner the value  $\gamma_{opt}$ for which $H_{\gamma_{opt}}$ leads to optimal search outcomes, i.e. $\pi_{w}^{\gamma_{opt}}(t)$ is maximal in the shortest time possible (see (\ref{eq:OptimalTimeGamma}) for the definition of $\gamma_{opt}$). In  \cite{ChildsGoldstone2004}, 
Childs and Goldstone elaborated on the interplay between the success probability (\ref{eq:SuccessProb}) and the overlap probabilities 

\begin{align}
\vert \langle s, \psi_0 \rangle \vert^2, \     \vert \langle e_w,\psi_0\rangle\vert^2, \  \vert \langle s, \psi_1 \rangle \vert^2,  \ \vert \langle e_w,\psi_1\rangle\vert^2
\end{align}
where $\psi_0$ (resp. $\psi_1$) refer to the ground (resp. first excited) state of $H_{\gamma}$. As such, we focus on a better understanding  of these overlap probabilities resulting in our first main contribution 
Theorem~\ref{prop:OverlapCloseE0AndE1}. This result provides conditions \eqref{eq:CGoverlapEpsilon} under which we can approximate 
and relate the $\psi_0$-eigenvalue $E_0$ (resp. $\psi_1$-eigenvalue $E_1$) with the square root of the graph's volume, i.e.
\begin{align}
\label{eq:EigenvaluesVolumeG}
E_0 \approx -\frac{\sqrt{ \langle \delta_w, \delta_w \rangle}}{\sqrt{vol(G)}}, \quad  E_1 \approx \frac{\sqrt{ \langle \delta_w, \delta_w \rangle}}{\sqrt{vol(G)}}.
\end{align}
We point out that for the complete graph on $N$ vertices the eigenvalues are given by 
\begin{align}
\label{eq:EigenValForGammaE}
E_{0} = - \frac{1}{\sqrt{N}}, \quad E_{1} = \frac{1}{\sqrt{N}},
\end{align}
and the  sufficient conditions  Theorem~\ref{prop:OverlapCloseE0AndE1} are satisfied.  
Furthermore, in this case we have $\langle \delta_w, \delta_w \rangle=1$ for all $w \in V$ and $vol(G)=N$ is nothing else but the number of vertices. 

In practice, it might be difficult to verify  \eqref{eq:CGoverlapEpsilon}  for general graphs. Therefore we introduced the parameter $\gamma_E$ in \eqref{eq:DefOfParameters} for which  the corresponding success probability $\pi_{w}^{\gamma_E}(t)$ takes the simple form \eqref{eq:successProb}. In fact, for a complete graph on $N$ vertices  we have $\gamma_E=\gamma_{opt}$  holds for all $N$ and hence the corresponding optimal success probability is easily computed using  \eqref{eq:successProb} and given by
\begin{align}
\label{eq:SuccessProbCompleteGraph}
\pi_{w}^{\gamma_{opt}}(t)  =   \Big(\frac{N-1}{N} \Big)  \sin^2\Big(\frac{ t}{\sqrt{N}} \Big) + \frac{1}{N}.
\end{align}
These observations have led us to the second part of this work: Does the equality $\gamma_{opt}=\gamma_E$ hold for other graphs? Or more specifically, is it possible to construct a graph such that the following properties hold:
\begin{align}
\label{eq:DesiredProperties}
\begin{cases}
\text{ A graph with variable volume $vol(G)$.} \\
\text{ The optimal success probability is well approximated by (\ref{eq:successProb})}. \\
\text{ $E_1$ is well approximated by $\frac{\sqrt{ \langle \delta_w, \delta_w \rangle}}{\sqrt{vol(G)}}$.}
\end{cases}
\end{align}

In section \ref{sec:SHL} we introduce a hypercubic lattice as a cartesian product of directed path graphs. To keep the discussion simple, we assume that the path graph has four vertices and that the transition probabilities of a random walker between these vertices are given via a parameter $p$, see Figures \ref{fig:1Dimlevel1Lattice} and  \ref{fig:2Dimlevel1Lattice}. This parameter $p$ can be interpreted as quantifying the database homogeneity/non-homogeneity and will play the role of the tuning parameter of the graph volume $vol(G)$. Despite the simplicity of this model, we obtain interesting results when investigating the properties
(\ref{eq:DesiredProperties}).

Our work is part of a long term study of  mathematical physics on fractals and self-similar graphs \cite{KassoStrichartz2005,KassoStrichartz2007,KassoSaloffCosteTeplyaev2008,Akkermans2009ComplexDim,Akkermans2012ThermoPhoton,Akkermans2012SpatialLogPeriodic,Akkermans2020AC_circuits,Akkermans2013quantumFieldsFractals,Dunne2012HeatKernels,
AlonsoRuiz2016Hanoi,HinzMeinert2020}, in which novel features of quantum processes on fractals can be associated with the unusual spectral and geometric properties of fractals compared to regular graphs and smooth manifolds.

\section{Continuous-time quantum walk on finite graphs}\label{sec:GroverFiniteGraph}

We start this section with some preliminary results that will be needed later for the proof of Proposition \ref{prop:OverlapProb} and Theorem \ref{prop:OverlapCloseE0AndE1}. Let $E_a$ and $\psi_a$ denote the eigenvalues and eigenvectors of $H_{\gamma}$, respectively. We assume that $\{\psi_a \}_{E_a \in \sigma(H_{\gamma})}$ is an orthonormal basis and in this notation,  $E_0$ and $\psi_0$ refer to the ground state, $E_1$ and $\psi_1$ refer to the first excited state, and so on. 
For ease of discussion, we will assume in this work that $E_0$ and $E_1$ are non-degenerate. For  $\gamma>0$ and $z \in \rho(\gamma\Delta_{G})$, the resolvent set of $\gamma\Delta_{G}$, we consider the following Green function
\begin{align}
\label{eq:GreenFct}
G_{\gamma}(z,w,w) :=\langle e_w,(\gamma\Delta_{G}-z)^{-1} e_w \rangle.
\end{align}
Let $\{   \phi_{\lambda} \ \vert \ \lambda \in \sigma(\Delta_{G}) \ \}$ be an orthonormal basis of eigenvectors of $\Delta_{G}$, and write 
\begin{align}
\label{eq:FourierCoeffi}
e_{w} = \sum_{\lambda \in \sigma(\Delta_{G})} a_{w,\lambda} \phi_{\lambda},  \quad \quad a_{w,\lambda} = \langle \phi_{\lambda},e_{w} \rangle.   
\end{align}
where the sum takes the eigenvalue multiplicities into account. The following result whose proof we omit, is elementary. 
\begin{lemma} 
\label{lem:somePropertiesGreen}
The following statements hold.
\begin{enumerate}
\item If $E_a \notin \sigma(\gamma \Delta_{G})$, then $\langle e_w,\psi_a \rangle  \neq 0$.
\item If $E_a \notin \sigma(\gamma \Delta_{G})$, then $G_{\gamma}(E_a,w,w) =1$.
\item If $E_a \notin \sigma(\gamma \Delta_{G})$, then $\langle e_w, ( \gamma \Delta_{G}  -  E_a )^{-1} \psi_a \rangle = \frac{1}{\langle \psi_a,e_w \rangle}$.
\item If $z \in \rho(\gamma\Delta_{G})$, then 
$$\begin{cases}
G_{\gamma}(z,w,w) = \sum_{\lambda \in \sigma(\Delta_{G})} \frac{\vert a_{w,\lambda}\vert^2}{\gamma\lambda - z} \\
G_{\gamma}'(z,w,w) = \frac{d}{dz} G_{\gamma}(z,w,w) =  \sum_{\lambda \in \sigma(\Delta_{G})} \frac{\vert a_{w,\lambda}\vert^2}{(\gamma\lambda - z)^2} = \langle e_w,(\gamma\Delta_{G}-z)^{-2}  e_w \rangle.
\end{cases}$$
\end{enumerate}
\end{lemma}

Next, for general finite graphs we derive formulas for the overlap probabilities.

\begin{proposition}[Overlap probabilities]
\label{prop:OverlapProb}
 Suppose that  $E_a \notin \sigma(\gamma \Delta_{G})$. Then we have
\begin{enumerate}
\item  $\vert \langle e_w,\psi_a\rangle\vert^2 =\frac{1}{G_{\gamma}'(E_a,w,w)}$.
\item $\vert \langle s, \psi_a \rangle \vert^2 =  \frac{ \langle \delta_w, \delta_w \rangle }{vol(G) E_a^2 G_{\gamma}'(E_a,w,w)}$.
\end{enumerate}
\end{proposition}
\begin{proof} The proof uses Lemma \ref{lem:somePropertiesGreen}. 
\begin{enumerate}
\item  Given that  $\psi_a = \langle e_w, \psi_a \rangle (\gamma \Delta_{G}  -  E_a )^{-1} e_w$, we have  
\
$$1=\langle \psi_a, \psi_a\rangle=\overline{\langle e_w, \psi_a \rangle }   \langle e_w,\psi_a \rangle     \langle e_w,(\gamma \Delta_{G}  -  E_a )^{-2} e_w \rangle=\langle e_w,(\gamma \Delta_{G}  -  E_a )^{-2} e_w \rangle  \vert \langle e_w,\psi_a \rangle \vert^2$$
The statement follows by Lemma \ref{lem:somePropertiesGreen}(4).
\item  Recalling that  $\Delta_{G}s=0$, we see that 
$$
-  E_a  \langle s,\psi_a \rangle = \langle s, ( \gamma \Delta_{G}  -  E_a ) \psi_a \rangle =  \langle s,V_w\psi_a \rangle = \langle s,e_w \rangle \langle e_w,\psi_a \rangle.
$$
Hence $\vert \langle s,\psi_a \rangle \vert^2 =   \frac{\vert \langle s,e_w \rangle \vert^2 \ \vert \langle e_w,\psi_a \rangle \vert^2}{E_a^2}$ and the result follows by $\vert \langle s,e_w\rangle \vert^2 = \frac{ \langle \delta_w, \delta_w \rangle }{vol(G)}$ and part (1).
\end{enumerate}
\end{proof}

To study the question of which parameter $\gamma$ the Hamiltonian $H_{\gamma}$ leads to optimal search outcomes, we will consider the following parameters with the assumption that each of the sets below is non-empty
\begin{align}
\label{eq:DefOfParameters}
\begin{cases} 
\gamma_s := \inf_{ \gamma \in (0, \infty)}  \big\{ \gamma  \ \big\vert \ \text{ such that } \vert \langle s , \psi_0 \rangle \vert^2 = \vert \langle s , \psi_1 \rangle \vert^2  \big\} \\
\gamma_w := \inf_{  \gamma \in (0, \infty)}  \big\{ \gamma  \ \big\vert \ \text{ such that } \vert \langle e_w , \psi_0 \rangle \vert^2 = \vert \langle e_w , \psi_1 \rangle \vert^2 \big\}. \\
\gamma_{E}:= \inf_{  \gamma \in (0, \infty)}  \big\{ \gamma  \ \big\vert \ \text{ such that } E_0=-E_1 \big\}.
\end{cases}
\end{align}

The following theorem establishes a relationship between the overlap probabilities and the eigenvalues $E_0$, $E_1$ and provides sufficient conditions to approximate these eigenvalues by the square root of the graph's volume.

\begin{theorem}
\label{prop:OverlapCloseE0AndE1}
Assume that there exist  $\gamma\in (0, \infty)$ and $\epsilon >0$ such that
\begin{align}
\label{eq:CGoverlapEpsilon}
\Big\vert \vert \langle s , \psi_0 \rangle \vert^2 - \vert \langle e_w , \psi_0 \rangle \vert^2 \Big\vert \leq \epsilon.
\end{align}
Then $$\Big\vert E_0^2 - \frac{\langle \delta_w, \delta_w \rangle}{vol(G)} \Big\vert \leq \epsilon.$$ 

Similarly, if the inequality (\ref{eq:CGoverlapEpsilon}) holds for $\psi_1$, then
\begin{align*}
\Big\vert E_1^2 - \frac{\langle \delta_w, \delta_w \rangle}{vol(G)} \Big\vert  \leq \Big( 1 + \frac{\langle \delta_w, \delta_w \rangle}{vol(G)} \frac{\vert  \vert \langle s , \psi_1 \rangle \vert^2 - \vert \langle s , \psi_0 \rangle \vert^2 \vert}{\vert \langle s , \psi_1 \rangle \vert^2  \vert \langle s , \psi_0 \rangle \vert^2} \Big) \epsilon
\end{align*}
\end{theorem}

\begin{proof} By Proposition \ref{prop:OverlapProb}(2), we have
\begin{align}
\Big\vert E_a^2 - \frac{\langle \delta_w, \delta_w \rangle}{vol(G)} \Big\vert 
&=\frac{\langle \delta_w, \delta_w \rangle}{vol(G)} \Big\vert  \frac{ 1- G_{\gamma}'(E_a,w,w)\vert \langle s, \psi_a \rangle \vert^2}{  G_{\gamma}'(E_a,w,w)\vert \langle s, \psi_a \rangle \vert^2} \Big\vert \\
&=\frac{\langle \delta_w, \delta_w \rangle}{vol(G)} \Big\vert  \frac{ \vert \langle e_w,\psi_a\rangle\vert^2 - \vert \langle s, \psi_a \rangle \vert^2}{  \vert \langle s, \psi_a \rangle \vert^2} \Big\vert 
\\
&=E_a^2 G_{\gamma}'(E_a,w,w) \Big\vert  \vert \langle e_w,\psi_a\rangle\vert^2 - \vert \langle s, \psi_a \rangle \vert^2 \Big\vert 
\end{align}
where in the second equality, we used $1=G_{\gamma}'(E_a,w,w)\vert \langle e_w,\psi_a\rangle\vert^2 $ (see Proposition \ref{prop:OverlapProb}(1)) and in the last equality, we used Proposition \ref{prop:OverlapProb}(2). The result follows from the following computations. First, we note that  Lemma \ref{lem:somePropertiesGreen}(4), $E_0 <0$ and $\sigma(\Delta_{G}) \subset [0,2]$ show
\begin{align}
\label{eq:ApproxE02Gprime}
E_0^2 G_{\gamma}'(E_0,w,w) &=E_0^2 \sum_{\lambda \in \sigma(\Delta_{G})} \frac{\vert a_{w,\lambda}\vert^2}{(\gamma\lambda + \vert E_0 \vert )^2}\\
\label{eq:ApproxE02Gprime}
& \leq \sum_{\lambda \in \sigma(\Delta_{G})} \vert  \langle \phi_{\lambda} ,e_{w} \rangle\vert^2 = \vert \vert e_w\vert \vert^2 = 1.
\end{align}
Moreover, Proposition \ref{prop:OverlapProb} (2) shows
\begin{align}
\Big\vert E_1^2 G_{\gamma}'(E_1,w,w)& - E_0^2 G_{\gamma}'(E_0,w,w) \Big\vert =  \frac{\langle \delta_w, \delta_w \rangle}{vol(G)} \frac{\Big\vert  \vert \langle s , \psi_1 \rangle \vert^2 - \vert \langle s , \psi_0 \rangle \vert^2 \Big\vert}{\vert \langle s , \psi_1 \rangle \vert^2  \vert \langle s , \psi_0 \rangle \vert^2}
\end{align}
\end{proof}
Complete graphs are examples for which the hypotheses of Theorem \ref{prop:OverlapCloseE0AndE1} are satisfied. In fact, if we consider the probabilistic graph Laplacian of a complete graph of $N$ vertices
\begin{align}
\label{eq:ProbLapCompleteGraph}
\Delta_G = \left( \begin{matrix}1 & - \frac{1}{N-1} & \dots & \dots & - \frac{1}{N-1}\\
- \frac{1}{N-1} & 1 & - \frac{1}{N-1} & \dots & - \frac{1}{N-1}\\
\vdots & \ddots & \ddots & \ddots & \vdots \\ 
 \vdots & \ddots & - \frac{1}{N-1} & 1 & - \frac{1}{N-1}\\
  - \frac{1}{N-1} & \dots & \dots & - \frac{1}{N-1} & 1\end{matrix}\right),
\end{align}
and set $\gamma =  \frac{N-1}{N}$, then a direct computation of the overlap probabilities gives
\begin{align}
\begin{cases}
\label{eq:OverlapProbCompleteGraphsCon}
\ \vert \langle s, \psi_0 \rangle \vert^2 = \vert \langle e_w,\psi_0\rangle\vert^2 = \frac{\sqrt{N}+1}{2 \sqrt{N}}\\
 \ \vert \langle s, \psi_1 \rangle \vert^2 = \vert \langle e_w,\psi_1\rangle\vert^2 = \frac{\sqrt{N}-1}{2 \sqrt{N}}
\end{cases}
\end{align}
It follows that \eqref{eq:CGoverlapEpsilon} holds for any $\epsilon > 0$, the eigenvalues are given by \eqref{eq:EigenValForGammaE}, and that, by definition, we have
 $\gamma_E =  \frac{N-1}{N}$. We remark that the Hamiltonian in \cite{AgliariEtal2010,ChildsGoldstone2004} is defined using the graph Laplacian $D-A$, where $D$ (resp.  $A$) is the degree (resp. adjacency) matrix of the graph. By the regularity of complete graphs, the probabilistic graph Laplacian \eqref{eq:ProbLapCompleteGraph} coincides with the graph Laplacian up to a multiple constant, i.e. $(N-1)\Delta=(D-A)$. Using either operators has no impact on the analysis since 
\begin{align*}
 \gamma \Delta - V_w =  \tilde{\gamma} (D-A) - V_w
\end{align*}
where $\gamma = (N-1)\tilde{\gamma}$. In particular, $\gamma = \gamma_E =  \frac{N-1}{N}$ if and only if $\tilde{\gamma}=\frac{1}{N}$, in which case, as proved  in \cite{ChildsGoldstone2004}, a quantum search on complete graphs recovers the optimal quadratic speedup.
Therefore, for complete graphs, Theorem \ref{prop:OverlapCloseE0AndE1} implies that $E_0=-E_1$.
The following result gives the consequences of assuming that $E_0=-E_1$ for a given graph with a fixed topology.


\begin{proposition}
\label{prop:SuccessProbability}
Suppose that there exists $\gamma \in (0,\infty)$ such that $E_0=-E_1$, then
\begin{align}
 \frac{\vert \langle e_w,\psi_1\rangle\vert^2}{\vert \langle s,\psi_1\rangle\vert^2} = \frac{\vert \langle e_w,\psi_0\rangle\vert^2}{\vert \langle s,\psi_0\rangle\vert^2}.
\end{align}
Consequently, we have  $   \overline{\langle \psi_1,s\rangle} \langle e_w, \psi_0 \rangle =  - e^{i 2 \theta} \overline{ \langle\psi_0 ,s \rangle  } \langle e_w, \psi_1  \rangle $ for some phase $ \theta \in [0,\pi)$. Moreover, the success probability reduces to 
\begin{align}
\label{eq:successProb}
\pi_{w}^{\gamma}(t)  =   4 \vert \langle s, \psi_0  \rangle \vert^2 \ \vert \langle e_w, \psi_1  \rangle \vert^2  \sin^2(E_1 t + \theta) + C + R(t)
\end{align}
where $C$, and $R(t)$ are given by
\begin{align}
\begin{cases}
C : = \vert \langle e_w, \psi_0  \rangle \vert^2 \vert \langle s, \psi_0  \rangle \vert^2 +  \vert \langle e_w, \psi_1  \rangle \vert^2 \vert \langle s, \psi_1  \rangle \vert^2 - 2 \vert \langle s, \psi_0  \rangle \vert^2 \ \vert \langle e_w, \psi_1  \rangle \vert^2  \\
R(t) := 2 Re\big( A(t) \overline{ r(t) } \ \big) + \vert r(t) \vert^2
\end{cases}
\end{align}
with
\begin{align}
\begin{cases}
\label{eq:AtAndRt}
A(t) := \langle e_w, \psi_0  \rangle \langle\psi_0 ,s \rangle \exp(-iE_0t) +  \langle e_w, \psi_1  \rangle \langle\psi_1 ,s \rangle \exp(-iE_1t) \\
r(t) := \sum_{a \geq 2}    \langle e_w, \psi_a  \rangle \langle\psi_a ,s \rangle \exp(-iE_at)             
\end{cases}
\end{align}
\end{proposition}
\begin{proof}
The first result follows by Proposition \ref{prop:OverlapProb} (1) \& (2). To prove the second result, we use $\langle e_w, \exp(-itH_{\gamma})  s \rangle = A(t) + r(t)$ and compute
\begin{align*}
\pi_{w}^{\gamma}(t)  &= \vert \langle e_w, \psi_0  \rangle \vert^2 \vert \langle s, \psi_0  \rangle \vert^2 +  \vert \langle e_w, \psi_1  \rangle \vert^2 \vert \langle s, \psi_1  \rangle \vert^2  \\
& \quad + \langle e_w, \psi_0  \rangle \langle\psi_0 ,s \rangle  \overline{\langle e_w, \psi_1  \rangle } \ \overline{ \langle\psi_1 ,s \rangle } \exp(i(E_1-E_0)t)  \\
& \quad + \overline{ \langle e_w, \psi_0  \rangle } \ \overline{ \langle\psi_0 ,s \rangle}  \langle e_w, \psi_1  \rangle \langle\psi_1 ,s \rangle \exp(-i(E_1-E_0)t) 
\\& \quad + A(t) \overline{r(t)} + \overline{A(t)} r(t) + r(t) \overline{r(t)}
\end{align*}
Using the first result, we add the second and third term together
\begin{align*}
 - \vert \langle s, \psi_0  \rangle \vert^2 \ \vert \langle e_w, \psi_1  \rangle \vert^2  \exp(i2(E_1 t + \theta)) 
 - \vert \langle s, \psi_0 \rangle \vert^2 \ \vert \langle e_w, \psi_1  \rangle \vert^2  \exp(-i2(E_1 t + \theta)) \\ =  - 2 \vert \langle s, \psi_0  \rangle \vert^2 \ \vert \langle e_w, \psi_1  \rangle \vert^2  \cos(2(E_1 t + \theta))
 \\ =  4 \vert \langle s, \psi_0  \rangle \vert^2 \ \vert \langle e_w, \psi_1  \rangle \vert^2  \sin^2(E_1 t + \theta)   - 2 \vert \langle s, \psi_0  \rangle \vert^2 \ \vert \langle e_w, \psi_1  \rangle \vert^2 
\end{align*}
\end{proof}

\section{Hypercubic lattices}
\label{sec:SHL}

\begin{figure}[htp]
\centering
\begin{minipage}{0.45\textwidth}
\centering
        \includegraphics[width=1.1\linewidth]{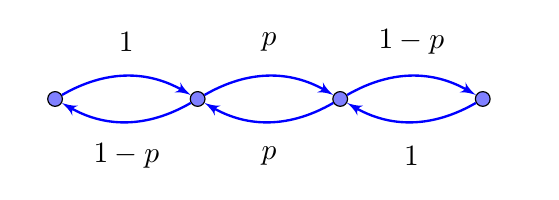}
\end{minipage}
\begin{minipage}{0.45\textwidth}
\centering
\begin{equation*}
\Delta_{G} =
\begin{pmatrix}
1 & -1 & 0 & 0   \\
p-1 & 1 & -p & 0  \\
0 & -p & 1 & p-1   \\
0 & 0 & -1 & 1
\end{pmatrix}. 
\end{equation*}
\end{minipage}
\vspace*{3mm}
\caption{The graph $G$ and the corresponding probabilistic graph Laplacian $\Delta_{G}$. When $p=1/2$, $\Delta_{G}$ becomes the standard probabilistic graph Laplacian. }
\label{fig:1Dimlevel1Lattice}
\end{figure}

In this section, we introduce a one-parameter family of Laplacians on the hypercubic lattice and investigate Grover's search algorithm numerically when using these Laplacians. Given a finite directed path graph $G=(V,E)$ with the vertices $V:=\{0,1,2,3\}$ and the edges $E: = \{(x,y) \in V \times V \ \vert \ \lvert x-y\rvert=1  \} $. We consider a random walk on $G$, where the transition probabilities $\{p(x,y)\}_{(x,y) \in V \times V }$ and the corresponding probabilistic graph Laplacian $\Delta_{G}$ are given for some $p \in (0,1)$ in Figure \ref{fig:1Dimlevel1Lattice}.  Note that $\Delta_{G}$ generates a random walk on $G$ with reflecting boundaries. This class of Laplacians was first investigated in \cite{TeplyaevSpectralZeta2007} and arises naturally when studying the unit-interval endowed with a particular fractal measure. For more on this Laplacian and some related work we refer to  \cite{Mograby2022AlmostMathieu,ChenTeplyaev2016,DerfelGrabnerVogl2012, NgaiTeplyaev2015, BirdNgaiTeplyaev2003,BaluMogOkoTep2021spectralAMO}. 
\begin{figure}[htp]
\centering
\centering
        \includegraphics[width=0.67\linewidth]{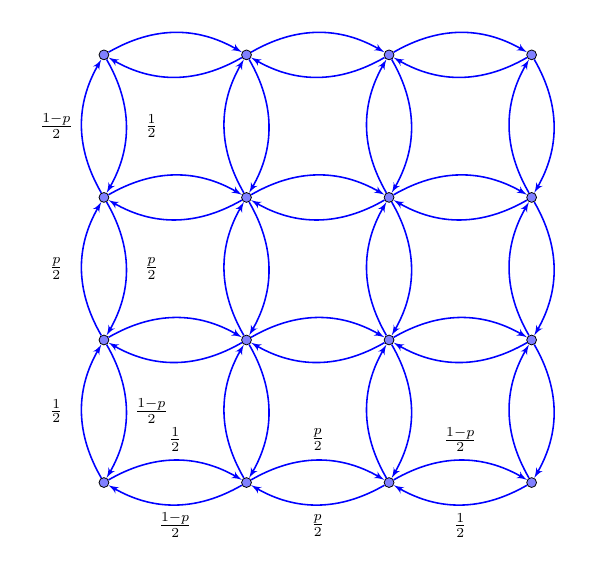}
\caption{The graph $G_2=G' \times G''$, where $G' =G'' = G$ and $G$ is the graph in Figure \ref{fig:1Dimlevel1Lattice}. The transition probabilities are given by $\Delta_{G_2}$.}
\label{fig:2Dimlevel1Lattice}
\end{figure}

We equip the vertices set $V$ with a measure, $\mu:V \to [0,\infty)$ and assume that it satisfies the Kolmogorov’s cycle condition \cite{DurrettBookProbability2019, GrigoryanIntroGraphs2018,KellerLenzBook2021,KellyBookReversibility2011}, i.e.
\begin{align}
\label{eq:measureDef}
           \mu(0)=1, \quad  \mu(x) = \mu(x-1) \frac{p(x-1,x)}{p(x,x-1)},    \ \  x \in V.
\end{align} 
One can easily verify that 
$\Delta_{G}$ is self-adjoint with respect to the inner product
\begin{equation}
\label{eq:InnerPro1D}
  \langle f,g \rangle_G =  \sum_{x \in V} \overline{f(x)} g(x) \mu(x),
\end{equation}
\\
where $f, g \in \mathcal{H}_G:=span\{ \ \delta_x \ \vert \ x \in V\}$. A \textit{$d$-dimensional hypercubic lattice} is constructed as the $d$-fold Cartesian product of finite directed path graphs $G_{d} = G' \times G'' \times .. $, where $G'=G''=\hdots=G$. For simplicity, we restrict our illustration to products of two graphs as the extension to higher dimensional products is straightforward.  We follow \cite{Strichartz2007Products} and denote by a prime anything having to do with the first graph, and by a double prime anything having to do with the
second graph. We recall that a Cartesian product of two graphs $G'=(V',E')$  and $G''=(V'',E'')$  is a graph $G_{2} = G' \times G''$ with the set of vertices $V(G_{2}) = \{(x', x'') \ \vert \ x' \in V', x'' \in V'' \ \}$,
where two vertices $\bar{x}=(x', x'')$ and $\bar{y}=(y', y'')$ are adjacent, i.e. $(\bar{x},\bar{y}) \in E(G_{2})$ if and only if $(x',y') \in E' \ \text{ and } x''=y''$ or $(x'',y'') \in E'' \ \text{ and } x'=y'$. We define a Laplacian on $G_2$ as a (normalized) Kronecker sum of $\Delta_{G'}$ and $\Delta_{G''}$, i.e.
\begin{align}
\Delta_{G_2} := \frac{1}{2}\big(\Delta_{G'} \otimes I + I \otimes \Delta_{G''}\big)
\end{align}
where $I$ is an identity matrix. An example of the random walk generated by $\Delta_{G_2}$ is illustrated in Figure \ref{fig:2Dimlevel1Lattice}. We associate each vertex $\bar{x}=(x', x'') \in V(G_{2}) $ with $\delta_{\bar{x}} := \delta_{x'} \otimes \delta_{x''}$, the tensor product $ \delta_{x'} $ and $ \delta_{x''}$. It follows that $\Delta_{G_2}$  is self-adjoint on the Hilbert space $\mathcal{H}_{G_2}=span\{ \ \delta_{\bar{x}} \vert \ \bar{x}=(x', x'') \in V(G_{2}) \ \}$ equipped with the inner product
\begin{equation}
\label{eq:InnerProhigherD}
  \langle f,g \rangle =  \sum_{\bar{x} \in V(G_2)} \overline{f(\bar{x})} g(\bar{x}) \mu_2(\bar{x}), \quad   \begin{cases}
  \   \langle \delta_{\bar{x}}, \delta_{\bar{y}} \rangle := \langle \delta_{x'},\delta_{y'} \rangle_{G'} \langle \delta_{x''},\delta_{y''}\rangle_{G''} \\
  \  \mu_2(\bar{x}):=\langle \delta_{\bar{x}}, \delta_{\bar{x}} \rangle
  \end{cases} 
\end{equation}


\subsection{Homogeneous versus non-homogeneous structures}

When $p=\frac{1}{2}$, then $\Delta_{G_d}$ recovers the standard probabilistic graph Laplacian as a generator of a symmetric random walk on $G_d$. In this case, it easy to compute that the symmetrizing measure \eqref{eq:InnerProhigherD} is the degree of the vertex, i.e.
\begin{align*}
\mu_d(\bar{x})=\langle \delta_{\bar{x}}, \delta_{\bar{x}} \rangle = \mu(x_1) \dots  \mu(x_d) = 2^d
\end{align*}
where $\bar{x}=(x_1,\dots, x_d) \in V(G_{d})$ is a non-boundary vertex. In particular, the measure is constant on the interior vertices, and as such, we say that  $G_d$  \textit{homogeneous}. On the other hand, when $p\neq \frac{1}{2}$, then $\Delta_{G_d}$ generates an asymmetric random walk on $G_d$. Consequently, the measure is vertex-dependent and varies on the interior vertices. In this case, we say $G_d$ is \textit{non-homogeneous}.    For an interpretation from a physics viewpoint and the relation to fractal media, the reader is referred to the introduction of \cite{ChenTeplyaev2016}.

\subsection{Numerical results}
\begin{figure}[!htb]
    \centering
    \begin{minipage}{.5\textwidth}
        \centering
        \includegraphics[width=1.11\linewidth]{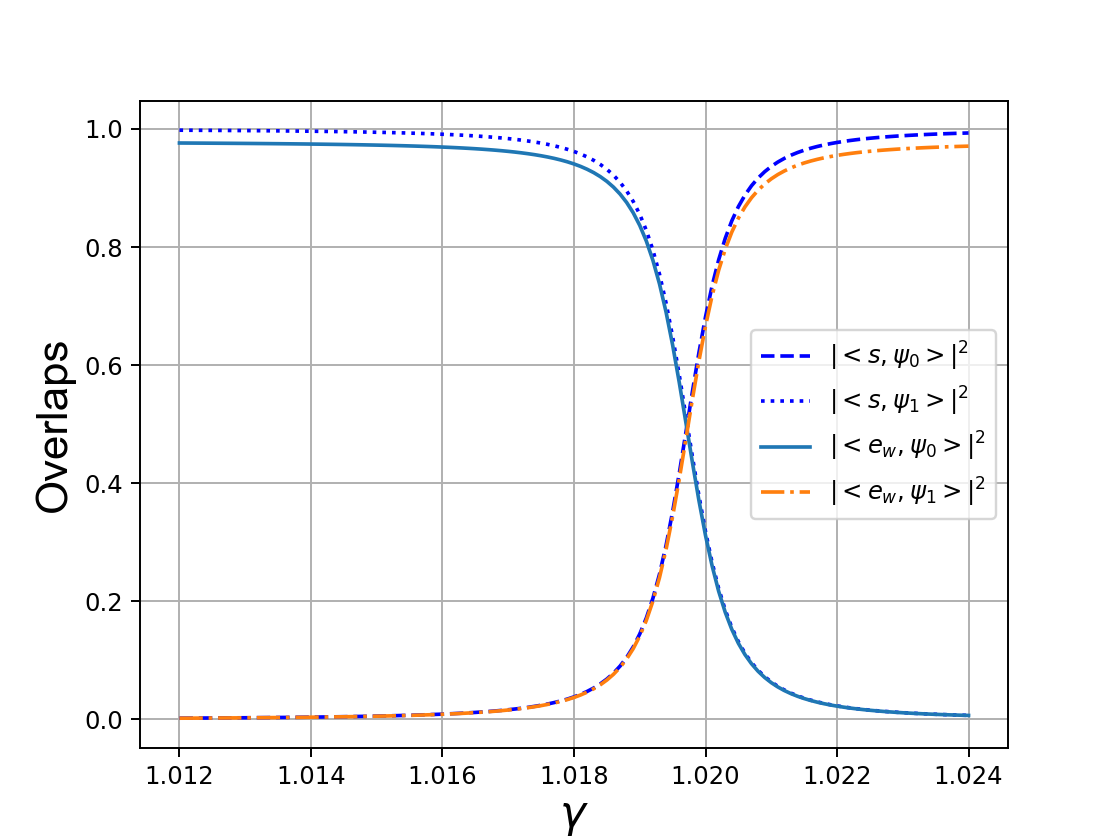}
    \end{minipage}%
    \begin{minipage}{0.5\textwidth}
        \centering
        \includegraphics[width=1.11\linewidth]{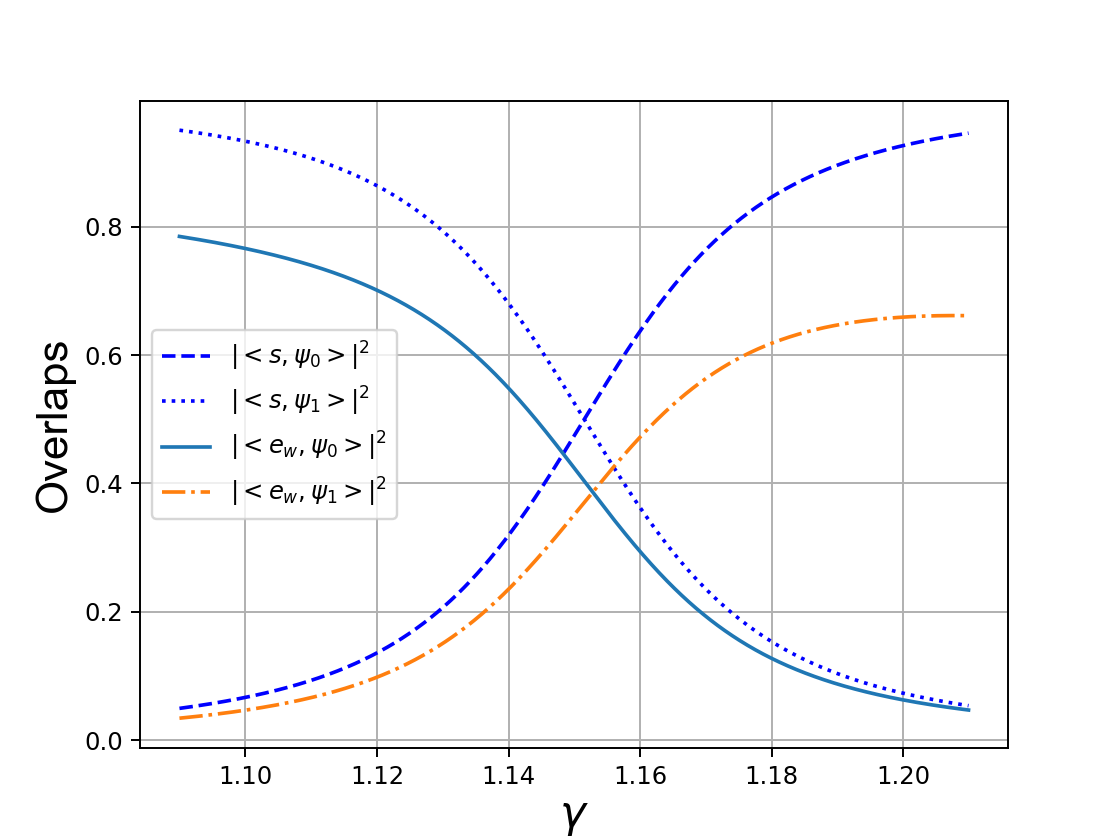}
    \end{minipage}
        \begin{minipage}{.5\textwidth}
        \centering
        \includegraphics[width=1.11\linewidth]{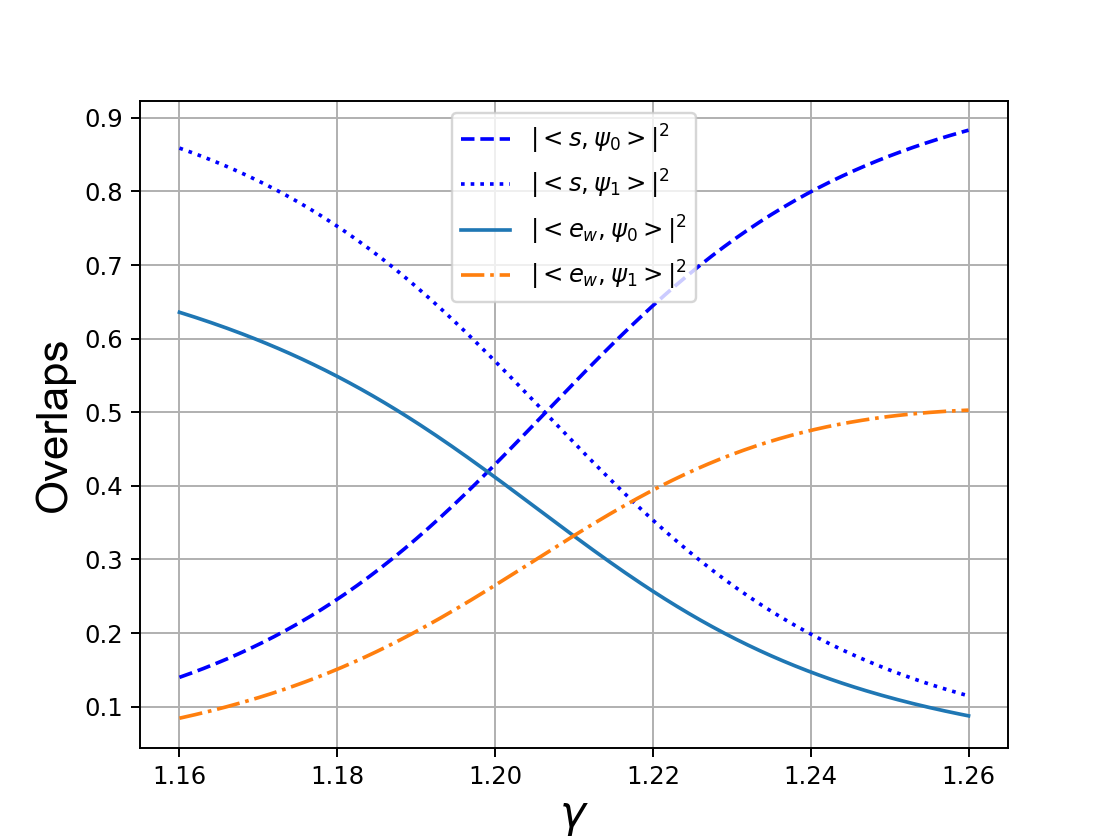}
    \end{minipage}%
    \begin{minipage}{0.5\textwidth}
        \centering
        \includegraphics[width=1.11\linewidth]{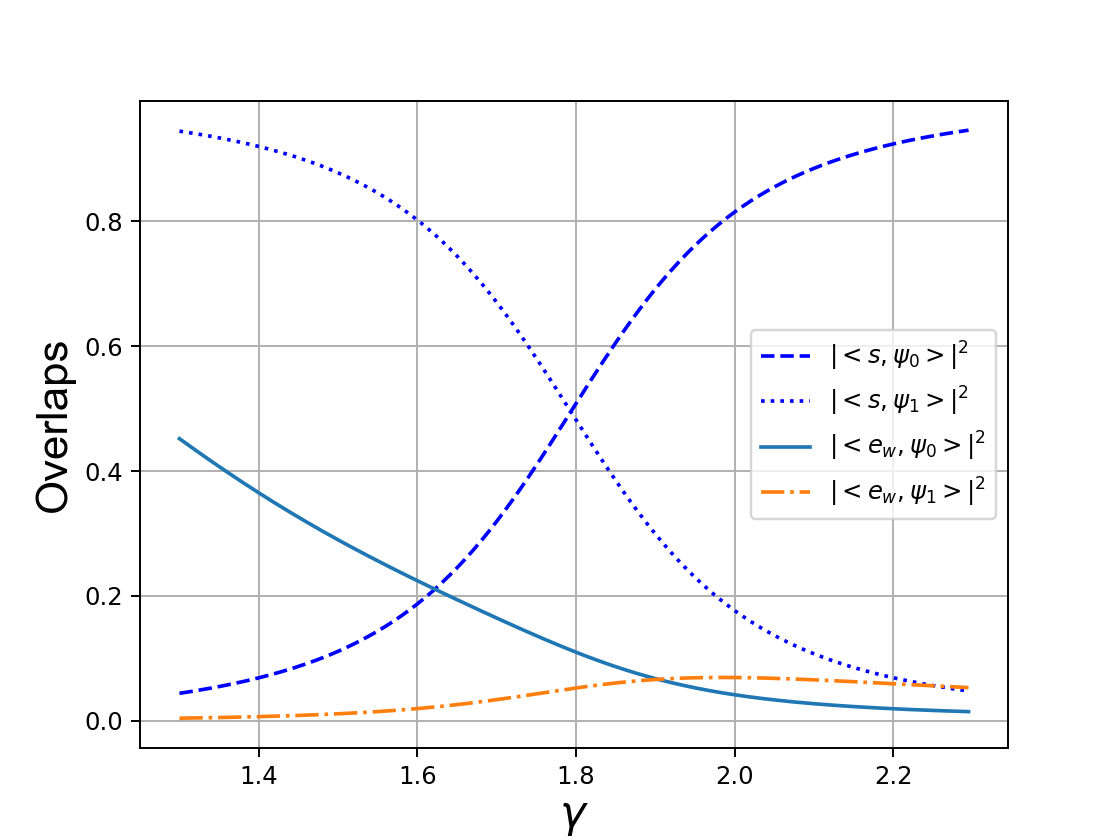}
    \end{minipage}
      \caption{Overlap probabilities as a function of $\gamma$ for $G_5$: (top, left) $p=0.91$, (top, right) $p=0.5$, (bottom, left) $p=0.4$, (bottom, right) $p=0.1$.}
    	  \label{fig:OverlapG5}
\end{figure}

In this section, we present some numerical results on the Grover's search algorithm on the graph $G_5$, for which the number of vertices is $\vert V(G_5)\vert = 1024$. The target vertex $w \in V(G_5)$ is assumed to be one of the corners of $G_5$. Our focus is to analyze the search algorithm based on the homogeneity versus non-homogeneity of the database, i.e., $p=1/2$ versus $p\neq 1/2$.  To this end, we plot  the overlap probabilities and the eigenvalues $E_0$, $E_1$ as functions of  $\gamma$, after which we determine and discuss their intersections at the points  $\gamma_s$,$\gamma_w$ and $\gamma_E$. 
Before proceeding, it is worth mentioning that for the complete graph of $N$ vertices, where $N\geq 2$, we have $\gamma_s \leq \gamma_E \leq \gamma_w$. When $N=2$ we have $\gamma_s=0$ and $\gamma_{w}={\infty}$, while for large $N$, we see that
\begin{align}
\gamma_s \approx     \gamma_E \approx  \gamma_w  \approx 1.
\end{align}
In the case of $G_5$, the plots of the overlap probabilities are depicted for several $p$ in Figure \ref{fig:OverlapG5}. The case $p=0.91$  ( top-left panel in Figure \ref{fig:OverlapG5})
is particularly interesting and is  qualitatively similar to the results for the complete graphs; see \cite[Figure 1]{ChildsGoldstone2004} for a comparison. We determine $\gamma_s$,$\gamma_w$ and $\gamma_E$ in Table \ref{tab:TableGammas}, and observe that in all cases
\begin{align*}
\gamma_s \leq \gamma_E \leq \gamma_w,
\end{align*}
where for larger $p$, $\gamma_E$ is increasingly squeezed between $\gamma_s$ and $\gamma_w$.
 
\begin{figure}[!htb]
    \centering
    \begin{minipage}{.5\textwidth}
        \centering
        \includegraphics[width=1.08\linewidth]{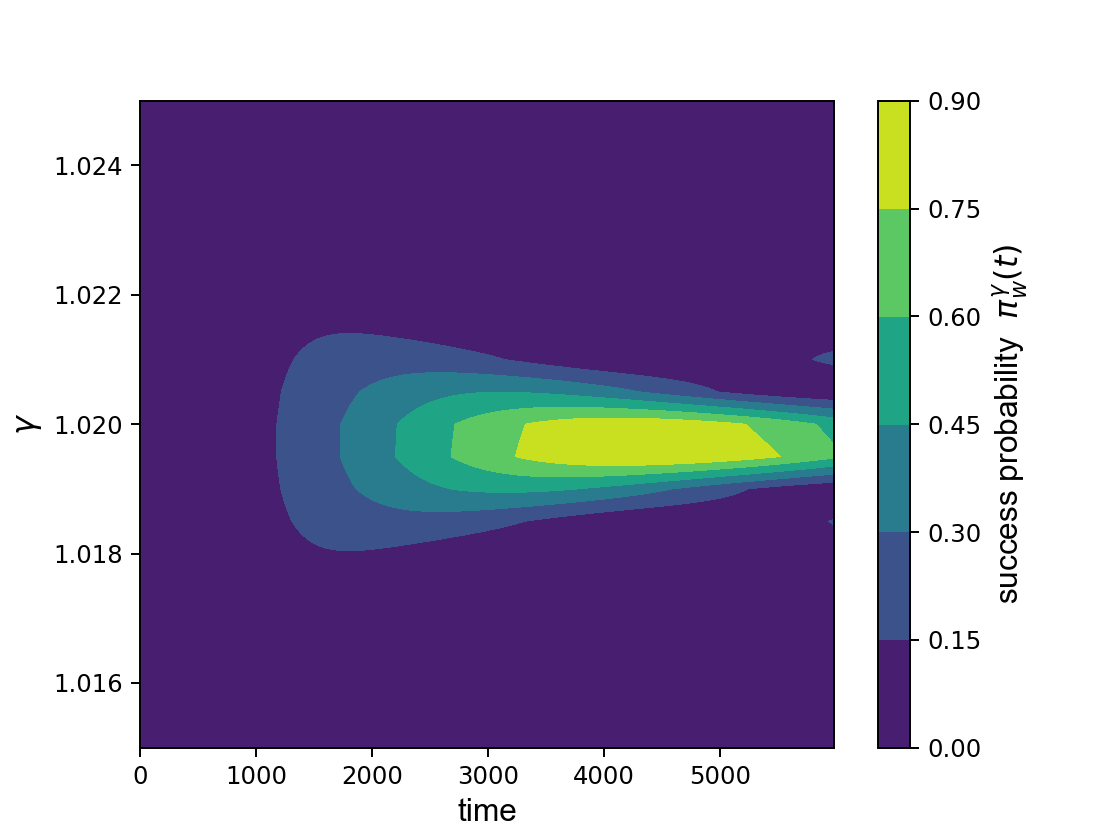}
    \end{minipage}%
    \begin{minipage}{0.5\textwidth}
        \centering
        \includegraphics[width=1.08\linewidth]{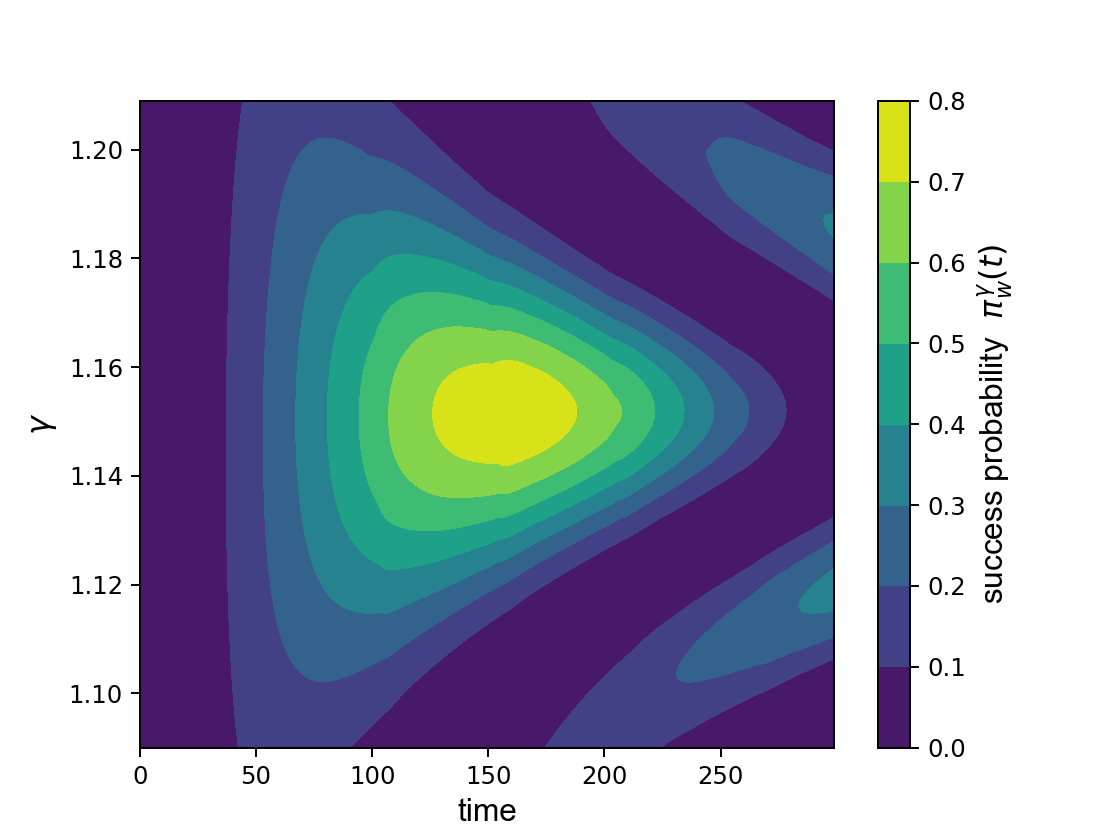}
    \end{minipage}
        \begin{minipage}{.5\textwidth}
        \centering
        \includegraphics[width=1.08\linewidth]{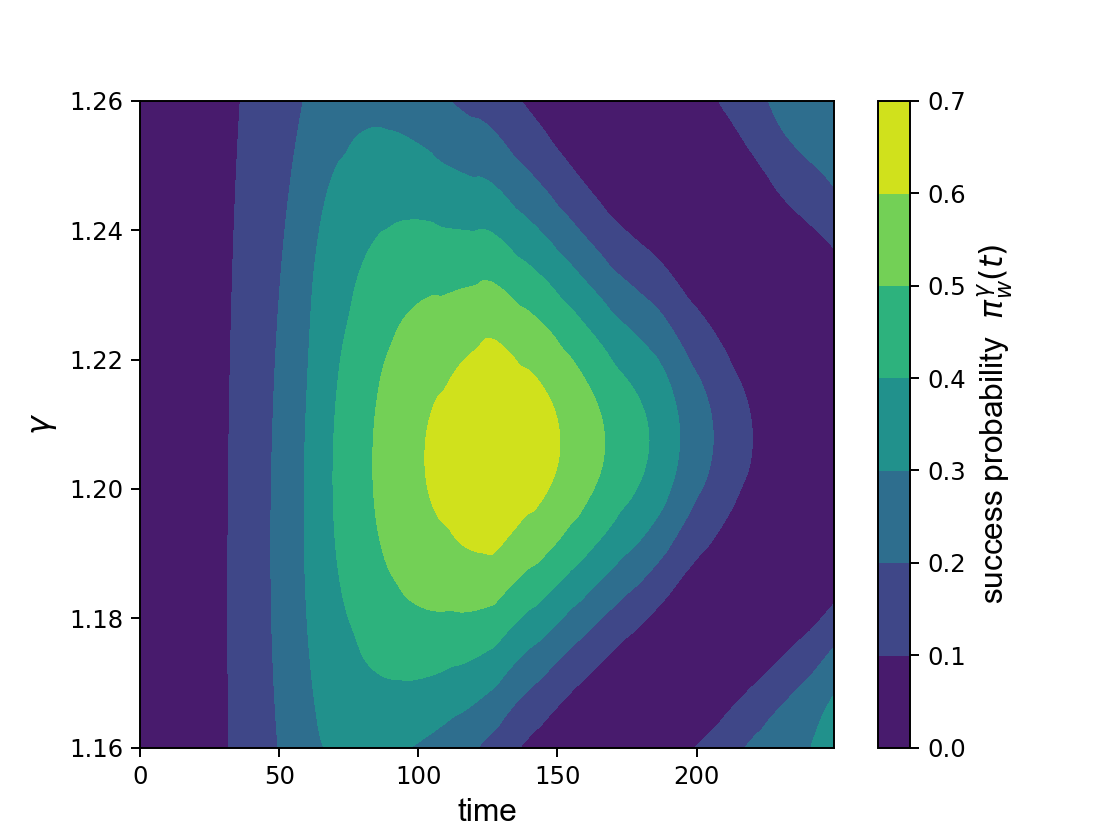}
    \end{minipage}%
    \begin{minipage}{0.5\textwidth}
        \centering
        \includegraphics[width=1.08\linewidth]{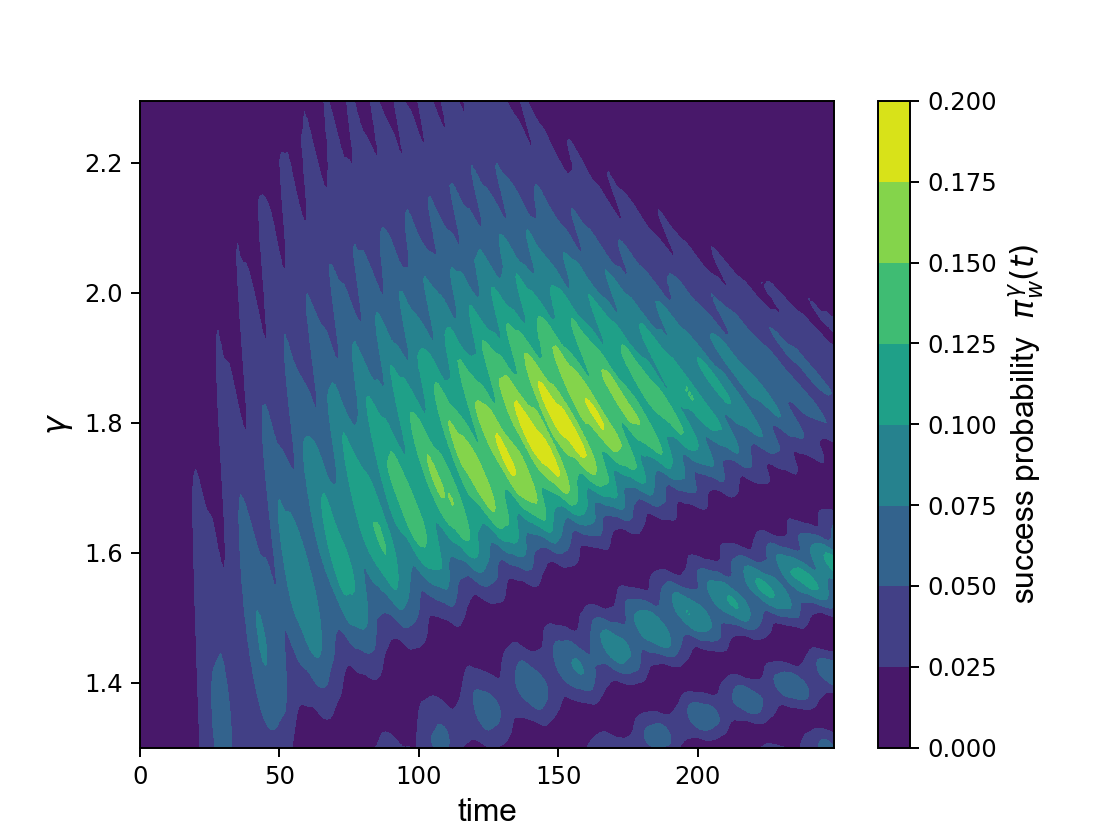}
    \end{minipage}
      \caption{Contour plot of the success probability $\pi_{w}^{\gamma}(t)$ as a function of the time $t$ and $\gamma$ for $G_5$: (top, left) $p=0.91$, (top, right) $p=0.5$, (bottom, left) $p=0.4$, (bottom, right) $p=0.1$.}
    	  \label{fig:timeEvolutionCompleteGraph}
\end{figure}
\begin{figure}[!htb]
    \centering
    \begin{minipage}{.5\textwidth}
        \centering
        \includegraphics[width=0.9\linewidth]{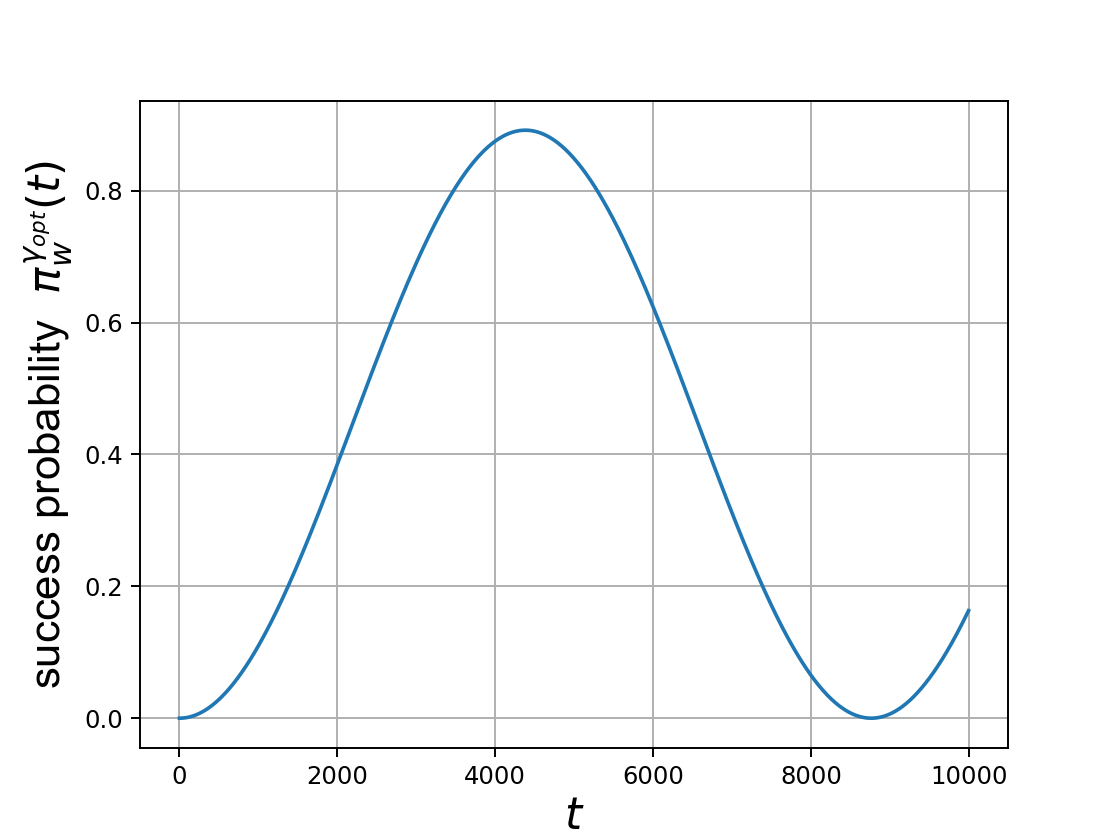}
    \end{minipage}%
    \begin{minipage}{0.5\textwidth}
        \centering
        \includegraphics[width=0.9\linewidth]{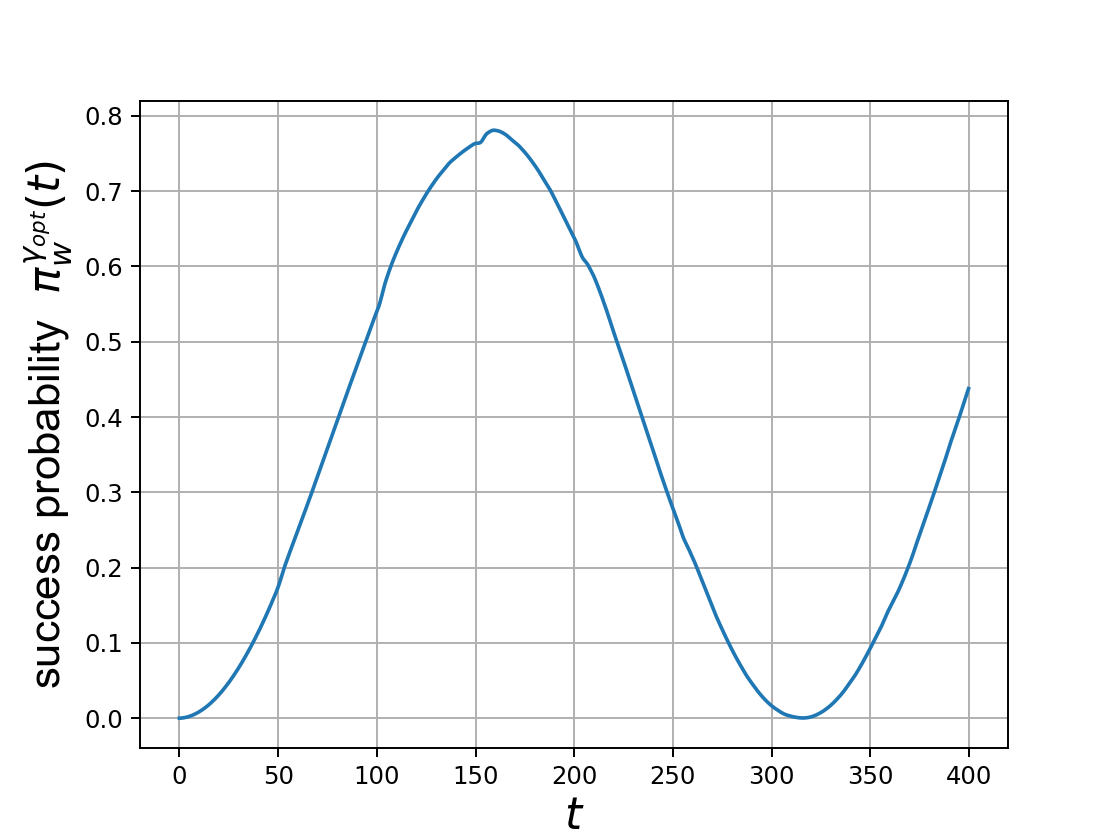}
    \end{minipage}
        \begin{minipage}{.5\textwidth}
        \centering
        \includegraphics[width=0.9\linewidth]{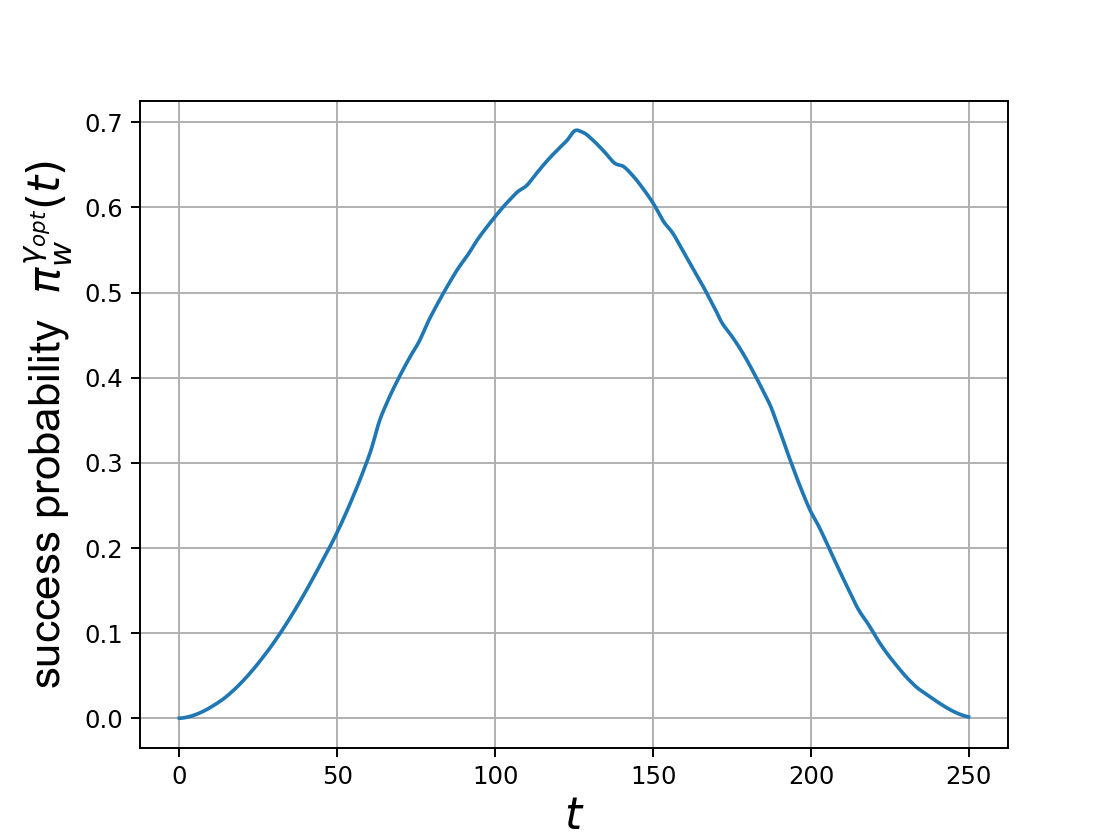}
    \end{minipage}%
    \begin{minipage}{0.5\textwidth}
        \centering
        \includegraphics[width=0.9\linewidth]{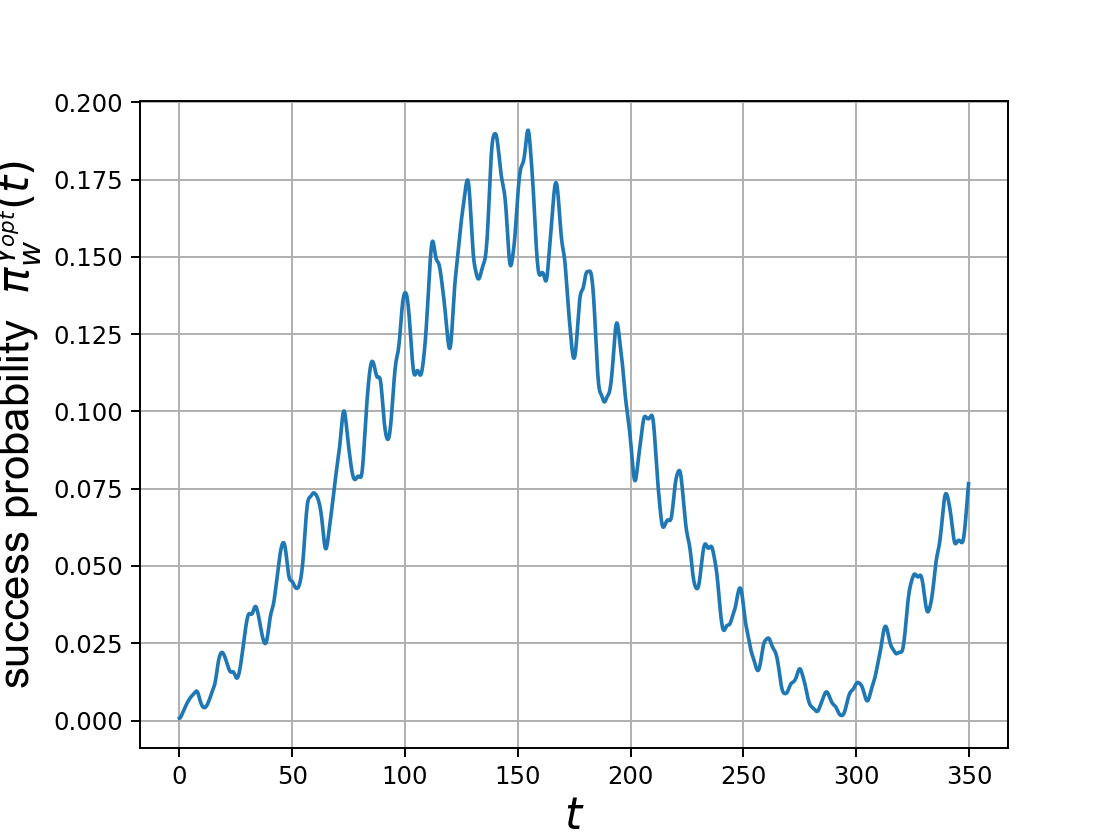}
    \end{minipage}
      \caption{Plot of the success probability $\pi_{w}^{\gamma}(t)$ as a function of time $t$ for $G_5$. We set $\gamma=\gamma_{opt}$ in all panels: (top, left) $p=0.91$, (top, right) $p=0.5$, (bottom, left) $p=0.4$, (bottom, right) $p=0.1$.}
    	  \label{fig:periodLattice}
\end{figure}

\begin{table}[htb!]
\centering
{\renewcommand{\arraystretch}{1.3}
 \begin{tabular}{c || c c   c || c} 
$p$ & $\gamma_s$ & $\gamma_w$ &$\gamma_E$ &  $\gamma_{opt}$ \\ [0.5ex] 
 \hline\hline 
 0.91 & 1.0197  & 1.0197 & 1.0197 &   1.0195      \\
 0.5 &  1.1515 & 1.1528 &  1.1521 &   1.1520   \\
0.4 &  1.2063 & 1.2099 &   1.2081 &   1.2061  \\
 0.1 &   1.7935 & 1.9035 & 1.8438 &  1.785  \\ [1ex] 
 \hline
 \end{tabular}
 }
 \caption{Numerical computation of $\gamma_s$, $\gamma_w$, $\gamma_E$ and $\gamma_{opt}$ for $G_5$ and different values of $p$.}
\label{tab:TableGammas}
\end{table}
\begin{table}[htb!]
\centering
{\renewcommand{\arraystretch}{1.3}
 \begin{tabular}{c || c c c || c c } 
$p$ & $E_0$ & $E_1$ & $\sqrt{\frac{\langle \delta_w, \delta_w \rangle}{vol(G_{5})}}$ & $t_{opt}$ & $\frac{\pi}{2}\sqrt{\frac{vol(G_{5})}{\langle \delta_w, \delta_w \rangle}}$ \\ [0.5ex] 
 \hline\hline 
 0.91 & -0.0004  & 0.0002 &    0.0003    & 4380  & 4535.8\\
 0.5 &  -0.010 & 0.0099 &     0.0113 & 159.4  & 138.52 \\
0.4 &  -0.0130 & 0.01189 &     0.0152 & 125.8 & 103.18 \\
 0.1 &   -0.0135 & 0.0085 &   0.0273 & 154.6 & 57.54 \\ [1ex] 
 \hline
 \end{tabular}
 }
 \caption{Numerical computation of $E_0$, $E_1$, $t_{opt}$ and $\frac{\pi}{2}\sqrt{\frac{vol(G_{5})}{\langle \delta_w, \delta_w \rangle}}$  for $G_5$ and different values of $p$.}
\label{tab:TableEigenvaluesLattice}
\end{table}
We also compute the success probability $\pi_{w}^{\gamma}(t)$ as a function of the time $t$ and $\gamma$, see Figure  \ref{fig:timeEvolutionCompleteGraph}. Then we determine for which values $(t,\gamma)$ the success probability $\pi_{w}^{\gamma}(t)$ is optimal, i.e.
\begin{align}
\label{eq:OptimalTimeGamma}
(t_{opt}, \gamma_{opt}) := \inf_{\gamma} \inf_{t} \big\{  (t,\gamma) \in [0,vol(G)] \times (0,\infty) : \pi_{w}^{\gamma}(t) \text{ attains abs. max.} \big\}
\end{align}
We observe that for  $p = 0.91$, we have $\gamma_{opt} \approx \gamma_{E} \approx \gamma_{s} \approx \gamma_{w}$, see Table \ref{tab:TableGammas}. (Note that for the complete graph of $N$ vertices, we have $\gamma_{opt} = \gamma_E $ for any $N \geq 2$.) Subsequently, we determine $E_0$ and $E_1$ as the eigenvalues of the ground- and first excited-state of  $H_{\gamma_{opt}}$. For comparison, we also compute $\sqrt{\frac{\langle \delta_w, \delta_w \rangle}{vol(G_{5})}}$ using
\begin{align*}
vol(G_{5}) = \Big(2 + \frac{2}{1-p} \Big)^5
\end{align*}
Note that $\langle \delta_w, \delta_w \rangle =1$ for $w$ located at one of the corners of $G_5$. Again, the results for $E_0$ and $E_1$ are in better agreement with $\sqrt{\frac{\langle \delta_w, \delta_w \rangle}{vol(G_{5})}}$ the larger we choose $p$.  We plot the success probability $\pi_{w}^{\gamma}(t)$ as a function of $t$, where we set $\gamma=\gamma_{opt}$, see Figure \ref{fig:periodLattice}. Concerning the observation that $\gamma_{opt} \approx \gamma_E$ for $p=0.91$, we note that the graph in the top-left panel in Figure \ref{fig:periodLattice} is in very good agreement with the analytical formula (\ref{eq:successProb}), i.e. 
\begin{align}
\label{eq:SuccessProbForP090}
\pi_{w}^{\gamma_{opt}}(t) \approx 0.89 \sin^2(E_1 t)
\end{align}
where the value of $E_1$ is given in Table \ref{tab:TableEigenvaluesLattice}. In contrast to the complete graphs where $\gamma_{opt} = \gamma_E $ holds for any number of vertices, we observe in $G_{5}$ that the deviation of $\gamma_E$ from $\gamma_{opt}$ increases for smaller values of $p$. In this case, formula (\ref{eq:successProb}) is less suitable for analyzing of $\pi_{w}^{\gamma_{opt}}(t)$. Indeed, Figure \ref{fig:periodLattice} (bottom-right panel) illustrates how $\pi_{w}^{\gamma_{opt}}(t)$ for $p=0.1$ exhibits a more irregular and oscillatory behavior.

Finally, we observe that $t_{opt}$ decreases with (large) $p$ and is comparable to $\frac{\pi}{2}\sqrt{\frac{vol(G_{5})}{\langle \delta_w, \delta_w \rangle}}$, see Table \ref{tab:TableEigenvaluesLattice}.
For example,  $\pi_{w}^{\gamma_{opt}}(t)$ for $p=0.91$ attains its maximum at $t_{opt} \approx 4380$, which is not practical for searching a database of $1024$ elements. On the other hand, Figure \ref{fig:GraphVolumeFuncP} seems to suggest  that choosing smaller $p$ might improve the optimal time of Grover's search algorithm. For example, when $p=0.4$, we observe a slight improvement from the homogeneous case $p=0.5$, see the corresponding $t_{opt}$ values in Table \ref{tab:TableEigenvaluesLattice}. However, it seems that this improvement in Grover's optimal time is lost when  $p$ get smaller, e.g.,  see the case $p=0.1$ in Table \ref{tab:TableEigenvaluesLattice}.

\section{Conclusions}

\begin{figure}[htp]
\centering
\centering
        \includegraphics[width=0.7\linewidth]{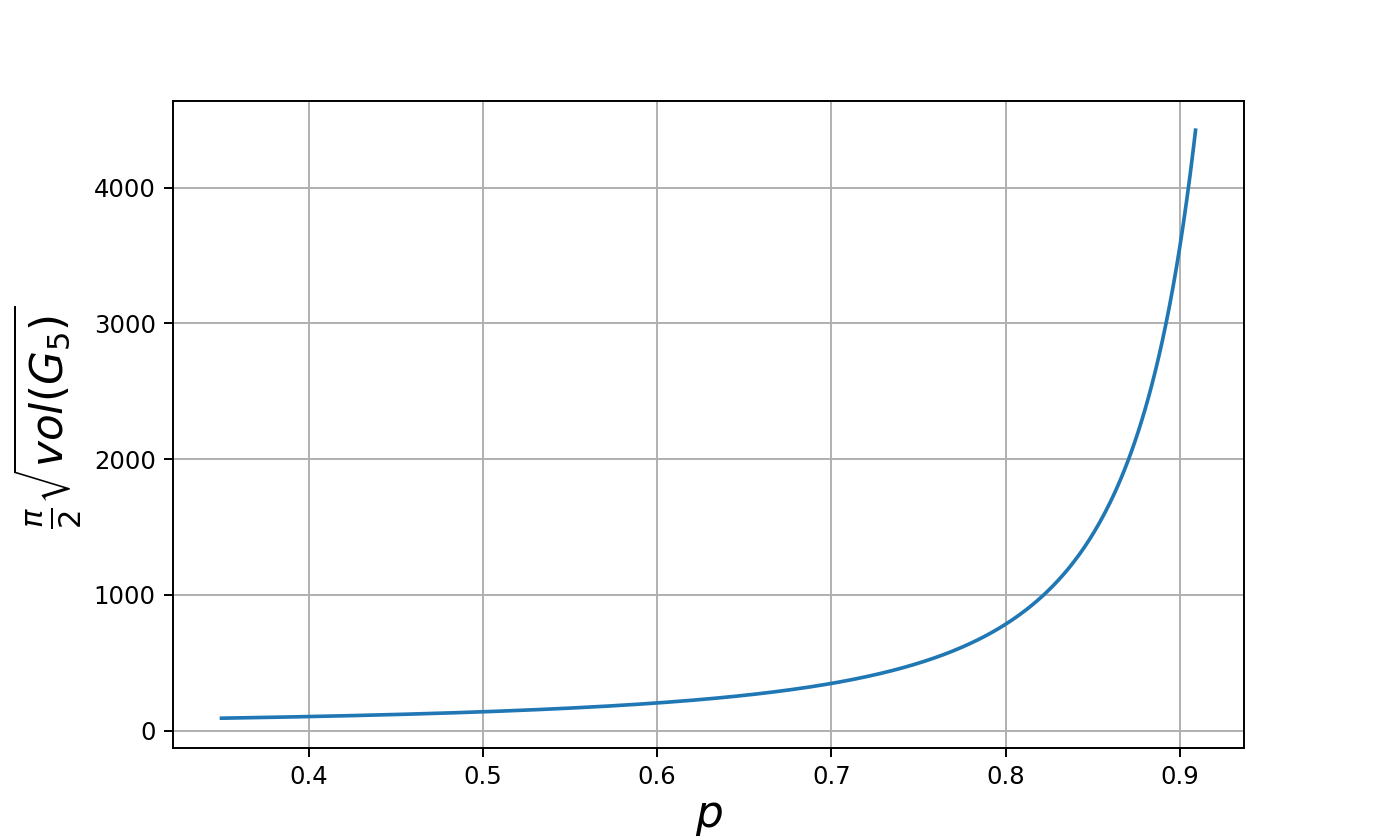}
\caption{The square root of the graph volume as a function of $p$ for $G_5$.}
\label{fig:GraphVolumeFuncP}
\end{figure}

In this paper, we discusse strategies for the application of Childs-Goldstone approach \cite{ChildsGoldstone2004} to Grover's quantum walk on graphs. A database is modeled by a finite (possibly directed) graph $G$ and the search algorithm is implemented using the family of Hamiltonians $\{H_{\gamma}\}_{\gamma \in (0,\infty)}$ in \eqref{eq:GroverHamilton}. Theorem \ref{prop:OverlapCloseE0AndE1} provides conditions \eqref{eq:CGoverlapEpsilon} on the overlaps probabilities that are sufficient to approximate and relate the eigenvalues $E_0$, $E_1 $ with the square root of the graph's volume. Complete graphs are examples for which the conditions \eqref{eq:CGoverlapEpsilon} hold for $\gamma =  \frac{N-1}{N}$ and any $\epsilon>0$, but this is not the case for the hypercubic lattices, see for instance the top-right panel in Figure \ref{fig:OverlapG5} (the homogeneous case $p=0.5$). On the other hand, we were able to tune the overlap probabilities on hypercubic lattices by inducing non-homogeneity measured by the parameter $p$. Indeed, the top-left panel in Figure \ref{fig:OverlapG5} evidences for $p=0.91$ the existence of $\gamma$ for which we have
\begin{align*}
\vert \langle s, \psi_0 \rangle \vert^2 \approx     \vert \langle e_w,\psi_0\rangle\vert^2 \approx   \vert \langle s, \psi_1 \rangle \vert^2  \approx \vert \langle e_w,\psi_1\rangle\vert^2.
\end{align*}
To deal with graphs for which it might be unfeasible to check \eqref{eq:CGoverlapEpsilon}, we introduced $\gamma_E$~ \eqref{eq:DefOfParameters} resulting in a simplified formula  for the corresponding success probability $\pi_{w}^{\gamma_E}(t)$~\eqref{eq:successProb}. In particular, to understand $\pi_{w}^{\gamma_E}(t)$, we need to analyze the following quantities:  $E_1$, $R(t)$ and the overlap probabilities
\begin{align*}
\vert \langle s, \psi_0 \rangle \vert^2, \     \vert \langle e_w,\psi_0\rangle\vert^2, \  \vert \langle s, \psi_1 \rangle \vert^2,  \ \vert \langle e_w,\psi_1\rangle\vert^2.
\end{align*}
This can be done rigorously for the complete graph of $N$ vertices, where $\gamma_E=\frac{N-1}{N}$ and $\pi_{w}^{\gamma_E}(t)$ is given by~\eqref{eq:SuccessProbCompleteGraph}. Furthermore, in this we have Complete graphs are particularly interesting as $\gamma_E=\gamma_{opt}$, i.e.
$H_{\gamma_E}$ leads to optimal search outcomes. We then ask whether equality $\gamma_{opt}=\gamma_E$ hold for other graphs, or, whether  we can construct a graph for which the properties \eqref{eq:DesiredProperties} hold. 

For this,  propose an approach for optimizing Grover's algorithm on hypercubic lattice graphs. In particular, 
in section \ref{sec:SHL}, we fix the graph topology (as hypercubic lattices) and instead vary the analysis structure on these graphs; this is done by varying the considered Laplacians. The main ideas of this approach are numerically demonstrated on hypercubic lattices. We restricted our investigation on $G_5$ and  focus on graph homogeneity/non-homogeneity effects on Grover's quantum walk. In upcoming work, we will extend our investigation to effects related to varying $d$ and the location of the target vertex $w$. 

In summary, we observe that the results for larger $p$ resemble those of the complete graphs qualitatively,  in particular $\gamma_{opt} \approx \gamma_E$ and the corresponding success probability is well approximated by \eqref{eq:successProb}. On the other hand, Grover's optimal times grow exponentially with $p$ in a pattern similar to the graph volume increase in Figure \ref{fig:GraphVolumeFuncP}. Choosing smaller $p$, like $p=0.4$, leads to a slight improvement in Grover's optimal time compared to the homogeneous case $p=0.5$. But this improvement doesn't continue with a further decrease of $p$ as the case $p=0.1$ in the bottom-right panel in Figure \ref{fig:periodLattice} shows.

\subsection*{Acknowledgments}

The work of G. Mograby and K. Okoudjou was supported by ARO grant W911NF1910366.  K. Okoudjou was additionally supported by NSF DMS-1814253. A.~Teplyaev was partially supported by NSF DMS grant 1950543 and by the Simons Foundation.

\bibliographystyle{plain}
\bibliography{BibList-lastUpdate-26Nov2021}

\end{document}